\documentclass[aps,prl,twocolumn,superscriptaddress,floatfix]{revtex4}
\usepackage[dvipdfmx]{graphicx}
\usepackage[dvipdfmx]{color}
\usepackage{amsmath}
\usepackage{amssymb}
\usepackage{dcolumn}
\usepackage{bm}
\usepackage{latexsym} 
\usepackage{setspace}
\usepackage{graphicx}
\usepackage{color}
\begin{document}
\title{Magnon polarons in the spin Peltier effect}
\author{Reimei Yahiro} 
\affiliation{Institute for Materials Research, Tohoku University, Sendai 980-8577, Japan} 
\author{Takashi Kikkawa}
\email{t.kikkawa@imr.tohoku.ac.jp}
\affiliation{Institute for Materials Research, Tohoku University, Sendai 980-8577, Japan}
\affiliation{WPI Advanced Institute for Materials Research, Tohoku University, Sendai 980-8577, Japan}
\author{Rafael Ramos}
\affiliation{WPI Advanced Institute for Materials Research, Tohoku University, Sendai 980-8577, Japan}
\author{Koichi Oyanagi} 
\affiliation{Institute for Materials Research, Tohoku University, Sendai 980-8577, Japan} 
\author{Tomosato Hioki} 
\affiliation{Institute for Materials Research, Tohoku University, Sendai 980-8577, Japan} 
\author{Shunsuke Daimon} 
\affiliation{Department of Applied Physics, The University of Tokyo, Tokyo 113-8656, Japan} 
\author{Eiji Saitoh}
\affiliation{Institute for Materials Research, Tohoku University, Sendai 980-8577, Japan}
\affiliation{WPI Advanced Institute for Materials Research, Tohoku University, Sendai 980-8577, Japan}
\affiliation{Department of Applied Physics, The University of Tokyo, Tokyo 113-8656, Japan}
\affiliation{Center for Spintronics Research Network, Tohoku University, Sendai 980-8577, Japan}
\affiliation{Advanced Science Research Center, Japan Atomic Energy Agency, Tokai 319-1195, Japan}
\date{\today}
\begin{abstract} 
We report the observation of anomalous peak structures induced by hybridized magnon-phonon excitation (magnon polarons) in the magnetic field dependence of the spin Peltier effect (SPE) in a Lu$_{2}$Bi$_{1}$Fe$_{4}$Ga$_{1}$O$_{12}$ (BiGa:LuIG) with Pt contact. 
The SPE peaks coincide with magnetic fields tuned to the threshold of magnon-polaron formation, consistent with the previous observation in the spin Seebeck effect. 
The enhancement of SPE is attributed to the lifetime increase in spin current caused by magnon-phonon hybridization in BiGa:LuIG. 
\end{abstract} 
\maketitle
%
%
%
\section{I.~~INTRODUCTION}
%
Magnetoelastic coupling (MEC), the interaction between spin waves (magnons) and lattice waves (phonons), was first investigated more than half a century ago \cite{Akhiezer1959SovPhysJETP,Kittel1958PhysRev,Kaganov1959SovPhysJETP} and has renewed attention in spintronics \cite{Dreher2012PRB,Ruckriegel2014PRB,Kamra2015PRB,Ogawa2015PNAS,Shen2015PRL,Takahashi2016PRL,Kikkawa2016PRL,Graczyk2017PRB,Flebus2017PRB,Cornelissen2017PRB,Bozhko2017PRL,Hashimoto2017NatCommun,Man2017PRB,Wang2018APL,Holanda2018NatPhys,Streib2018PRL,Schmidt2018PRB,Shan2018APL,Hayashi2018PRL,Shen2019PRB,Rameshti2019PRB,Simensen2019PRB,Sivarajah2019JAP,An2019,Ramos2019NatCommun,Thingstad2019PRL,Streib2019PRB}. 
By the MEC, magnons and phonons, in the vicinity of the crossings of their dispersion relations, are hybridized into quasiparticles called magnon polarons that share mixed magnonic and phononic characters \cite{Kamra2015PRB,Shen2015PRL,Kikkawa2016PRL,Flebus2017PRB,Cornelissen2017PRB,Schmidt2018PRB,Shan2018APL,Hayashi2018PRL,Shen2019PRB,Rameshti2019PRB,Simensen2019PRB,Thingstad2019PRL,Streib2019PRB,Sivarajah2019JAP,An2019,Ramos2019NatCommun}. 
Magnon polarons can convey spin information with velocities close to those of phonons, much faster than the magnon velocities in the dipolar magnon regime \cite{Ogawa2015PNAS,Shen2015PRL,Bozhko2017PRL,Hashimoto2017NatCommun}. 
Besides, thanks to the long-lived phononic constituent, magnon polarons may have longer lifetimes than pure magnons and can enhance the spin-current related phenomena, such as the spin Seebeck effect \cite{Kikkawa2016PRL,Flebus2017PRB,Cornelissen2017PRB,Wang2018APL,Schmidt2018PRB,Shan2018APL,Shen2019PRB,Ramos2019NatCommun} 
 and spin pumping \cite{Hayashi2018PRL}.  \par
%
\begin{figure}[htb]
\begin{center}
\includegraphics[width=8cm]{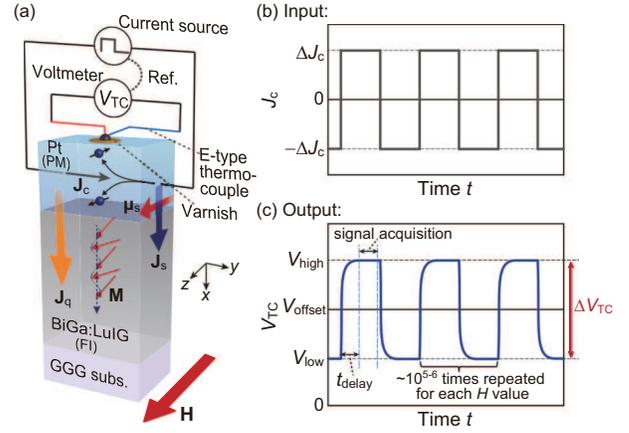}
\end{center}
\caption{(a) A schematic illustration of the SPE in the PM/FI junction (Pt/BiGa:LuIG/GGG sample) and measurement
system with a thermocouple (TC).  
The heat current ${\bf J}_{\rm q}$ induced by the SPE leads to a temperature change between the Pt and  BiGa:LuIG films in the vicinity of its interface, which is detected by the E-type TC attached on top of the Pt film with varnish \cite{Itoh2017PRB}. 
(b) Input signal: A square-wave charge current $J_{\rm c}$ with amplitude $\Delta J_{\rm c}$. 
(c) Output signal: A TC voltage $V_{\rm TC}$ that responds to the change in the $J_{\rm c}$ polarity, $\Delta V_{\rm TC} (\equiv V_{\rm high}-V_{\rm low})$, which is originated from the SPE-induced temperature modulation of the Pt ($\propto \Delta J_{\rm c}$) \cite{Itoh2017PRB}. Here, the Joule-heating-induced temperature modulation ($\propto \Delta J_{\rm c}^2$) is constant in time and can thus be excluded. 
}
\label{fig:1}
\end{figure}
%
%
\begin{figure}[htb]
\begin{center}
\includegraphics[width=85mm]{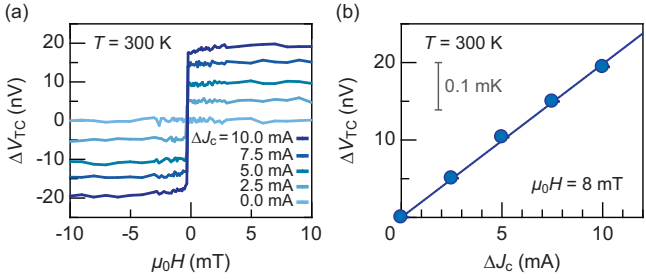}
\end{center}
\caption{
(a) $\Delta V_{\rm TC}\left( H \right)$ of the Pt/BiGa:LuIG/GGG sample for several $\Delta J_{\rm c}$ values measured at $T=300~\textrm{K}$ and $\left\vert \mu_0 H\right\vert <10~\textrm{mT}$.  
(b) $\Delta V_{\rm TC}\left( \Delta J_{\rm c} \right)$ measured at $T=300~\textrm{K}$ and $\mu_0 H = 8~\textrm{mT}$. 
The gray scale bar in (b) represents the temperature change $\Delta T$ of $0.1~\textrm{mK}$, estimated by the relation $\Delta T = \Delta V_{\rm TC}/S_{\rm TC}$.
}
\label{fig:2}
\end{figure}
%
%
%
%
The spin Seebeck effect (SSE) \cite{SSE_review} refers to the generation of a spin current (${\bf J}_{\rm s}$) as a result of a temperature gradient ($\nabla T$) in magnetic materials with metallic contacts. In the SSE, a magnon flow in a magnet is converted into a conduction-electron spin current at the magnet/metal interface and detected as a transverse electric voltage via the inverse spin Hall effect (ISHE) \cite{SHE_Hoffman,SHE_Sinova}.  
Recent experiments have revealed that small asymmetric peak structures appear in magnetic field-dependent (longitudinal) SSE voltages, which are interpreted as the enhanced spin current caused by the long-lived magnon-polaron formation  \cite{Kikkawa2016PRL,Flebus2017PRB,Cornelissen2017PRB,Wang2018APL,Schmidt2018PRB,Ramos2019NatCommun}  .
The peaks show up at magnetic fields $H$ at which the phonon dispersion curves become tangential to the magnon dispersion, i.e., when the magnon and phonon frequencies $\omega$, wave numbers $k$, and group velocities become the same. 
Under these ``touching'' conditions, the magnon and phonon modes can be coupled over the largest volume in $k$-space (see Fig. 2 of Ref. \cite{Flebus2017PRB}), such that the phase space portion over which the lifetimes of spin current are enhanced with respect to the uncoupled situation is maximal, leading to the enhancement of the SSE \cite{Kikkawa2016PRL,Flebus2017PRB}. \par 
In this study, we explore the magnon-polaron features in the spin Peltier effect (SPE), the reciprocal effect of the SSE, referring to the heat-current generation as a result of a spin current \cite{Flipse2014PRL,Basso2016PRB,Daimon2016NatCommun,Daimon2017PRB,Uchida2017PRB,Ohnuma2017PRB,Itoh2017PRB,Seki2018APL,Yagmur2018JPhysD,Sola2019SciRep,Daimon2019}. 
According to the Boltzmann transport theory \cite{Flebus2017PRB}, the bulk spin-transport coefficients for the SSE and SPE are linked by the Onsager reciprocal relation, so that similar small magnon-polaron peaks may show up also in the SPE. 
However, experimentally measured quantities are not the bulk transport coefficients, and thereby whether or not the hybridization peaks appear in the SPE is a nontrivial problem, which should be addressed experimentally. 
Nevertheless, there has been an issue to explore it; a measured SPE signal itself is usually very small. When measuring the SPE electrically by a thermocouple, the detected voltages are $\sim 10^{2-3}$ times smaller than the typical SSE electric voltages measured via the ISHE \cite{Flipse2014PRL,Itoh2017PRB,SSE_review}. 
Here, to overcome this issue, we use a Bi- and Ga-substituted lutetium iron garnet Lu$_{2}$Bi$_{1}$Fe$_{4}$Ga$_{1}$O$_{12}$ (BiGa:LuIG) film that exhibits large magnon-polaron peaks in SSE voltages [eight times greater than conventional Y$_{3}$Fe$_5$O$_{12}$ (YIG) films at room temperature, mainly due to its reduced magnon lifetime compared to YIG] \cite{Ramos2019NatCommun}.  
By careful measurements, magnon polarons turned out to manifest as peaks in magnetic field-dependent SPE signals, as with the SSE. 
Our results provide an important step towards a complete physical picture of magnon-polaron transport in magnetic insulators. \par 
%
%
%
\section{II.~~EXPERIMENTAL PROCEDURE} \label{sec:procedure}
%
%
%
\begin{figure}[htb]
\begin{center}
\includegraphics[width=8.4cm]{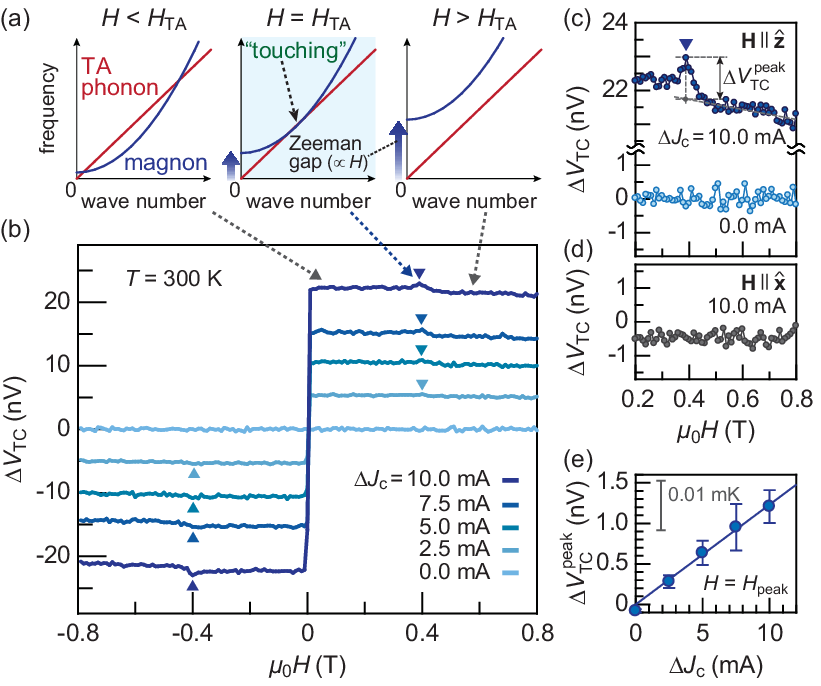}
\end{center}
\caption{
(a) Schematics of the magnon and TA-phonon dispersion relations for BiGa:LuIG when $H<H_{\rm TA}$, $H=H_{\rm TA}$, and $H>H_{\rm TA}$. 
At $T=300~\textrm{K}$ the touching frequency and wave number at $H=H_{\rm TA}$ are $\omega/2\pi \sim 0.02~\textrm{THz}$ and $k \sim 5 \times 10^7 ~\textrm{m}^{-1}$, respectively \cite{Ramos2019NatCommun}.
(b) $\Delta V_{\rm TC}\left( H \right)$ for several $\Delta J_{\rm c}$ values measured at $T=300~\textrm{K}$ and $\left\vert \mu_0 H\right\vert <0.8~\textrm{T}$. 
The $\Delta V_{\rm TC}$ peaks at $H_{\rm TA}$ are marked by blue triangles. 
(c) Magnified views of $\Delta V_{\rm TC}\left( H \right)$ around $H_{\rm TA}$ for $\Delta J_{\rm c} = 10.0$ and $0.0~\textrm{mA}$, where ${\bf H} \;||\; \mathbf{\hat{z}}$ (i.e., ${\bm \mu}_{\rm s} \;||\; {\bf M}$).
(d) Magnified view of $\Delta V_{\rm TC}\left( H \right)$ around $H_{\rm TA}$ for $\Delta J_{\rm c} = 10.0~\textrm{mA}$, where ${\bf H} \;||\; \mathbf{\hat{x}}$ (i.e., ${\bm \mu}_{\rm s} \perp {\bf M}$). 
(e) Peak amplitude $\Delta V_{\rm TC}^{\rm peak}$ at $H_{\rm TA}$ as a function of $\Delta J_{\rm c}$ at $T=300~\textrm{K}$. 
The gray scale bar in (e) represents $\Delta T = 0.01~\textrm{mK}$. 
}
\label{fig:3}
\end{figure} 
%
%
\begin{figure*}[htb]
\begin{center}
\includegraphics[width=13cm]{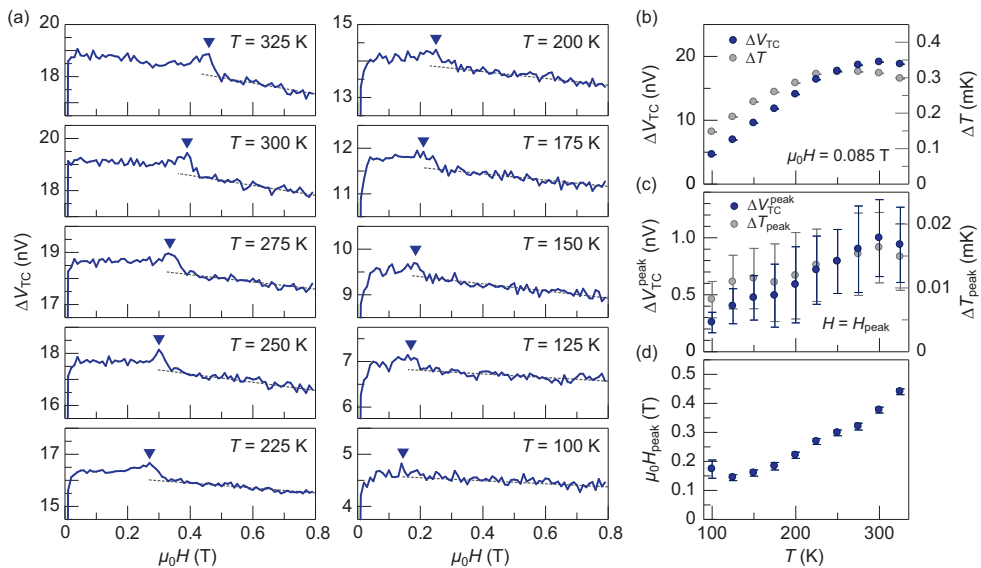}
\end{center}
\caption{
(a) $\Delta V_{\rm TC}\left( H \right)$ measured at various temperatures from $T =325$ to $100~\textrm{K}$ for $\Delta J_{\rm c}=10~\textrm{mA}$ and $\mu_0 H <0.8~\textrm{T}$. 
The $\Delta V_{\rm TC}$ peaks at $H_{\rm TA}$ are marked by blue triangles. 
The dashed straight lines are eye guides. 
(b) $T$ dependences of $\Delta V_{\rm TC}$ (blue plots, left vertical axis) and $\Delta T$ (gray plots, right vertical axis) 
at $\mu_0 H = 0.085~\textrm{T}$ ($<\mu_0 H_{\rm peak}$). 
(c) $T$ dependences of the peak amplitude $\Delta V_{\rm TC}^{\rm peak}$ (blue plots, left vertical axis) and $\Delta T_{\rm peak}$ (gray plots, right vertical axis) at the peak field $H_{\rm peak}$. 
(d) $T$ dependence of $H_{\rm peak}$. 
The $\Delta T$ values are estimated by the relation $\Delta T = \Delta V_{\rm TC}/S_{\rm TC}$, where $S_{\rm TC}$ is the $T$-dependent Seebeck coefficient of TC. 
The data for $250~\textrm{K} \leq T \leq 325~\textrm{K}$ ($100~\textrm{K} \leq T \leq 225~\textrm{K}$) was obtained by averaging 6-times (12-times) repetition measurements [that took $\sim 1$ day ($\sim 2$ days) for the data acquisition; see also Ref. \cite{polarity-change} and Sec. B of SM \cite{SM}].
}
\label{fig:4}
\end{figure*}
%
%
Figure \ref{fig:1}(a) shows a schematic of the SPE in a paramagnetic
metal (PM)/ferrimagnetic insulator (FI) junction, where the PM (FI) is Pt  (BiGa:LuIG) in the present study. 
The SPE appears as a result of a spin current induced by the spin Hall effect (SHE) in PM \cite{Flipse2014PRL,Daimon2016NatCommun,Daimon2017PRB,Itoh2017PRB}.  
When a charge current ${\bf J}_{\rm c} =J_{\rm c}\mathbf{\hat{y}}$ is applied to the PM, a spin current ${\bf J}_{\rm s}=J_{\rm s}\mathbf{\hat{x}}$ is generated and creates a nonequilibrium spin accumulation ${\bm \mu}_{\rm s}$ at the PM/FI interface \cite{SHE_Hoffman,SHE_Sinova,Flipse2014PRL,Daimon2016NatCommun,Daimon2017PRB,Itoh2017PRB}
\begin{equation}
{\bm \mu}_{\rm s} \; \propto \;  \theta_{\rm SHE}  {\bf J}_{\rm c} \times (-\mathbf{\hat{x}}) \; \propto \;  \theta_{\rm SHE} J_{\rm c} \mathbf{\hat{z}},
\label{eq:spin-accumulation}
\end{equation}
where $\theta_{\rm SHE}$ is the spin Hall angle of PM.  
By the interfacial spin-exchange coupling, ${\bm \mu}_{\rm s}$ causes a nonequilibrium magnon creation or annihilation scattering;  when ${\bm \mu}_{\rm s}$ is parallel (antiparallel) to the equilibrium magnetization (${\bf M}$) in FI, the number of magnons in FI is increased (decreased). 
This process accompanies a heat flow  ${\bf J}_{\rm q}$ between the PM/FI and thereby modulates the system temperature. The SPE-induced temperature modulation $\Delta T_{\rm SPE}$ satisfies the following relation \cite{Daimon2016NatCommun,Daimon2017PRB,Itoh2017PRB}
\begin{equation}
\Delta T_{\rm SPE} \; \propto \; {\bm \mu}_{\rm s} \cdot {\bf M} \; \propto \;  ({\bf J}_{\rm c} \times {\bf M}) \cdot \mathbf{\hat{x}}.
\label{eq:SPE}
\end{equation}
\par
We prepared a Pt($5~\textrm{nm}$)-strip/BiGa:LuIG($3~\mu \textrm{m}$) bilayer film, where the numbers in parentheses represent the thickness. 
The single-crystalline BiGa:LuIG (100) film was grown by liquid phase epitaxy on Gd$_{3}$Ga$_{5}$O$_{12}$ (GGG) (100) substrate (a $2 \times 1 \times 0.5~\textrm{mm}^3$ rectangular shape).
Before the Pt deposition, the BiGa:LuIG/GGG was cleaned with acetone in an ultrasonic bath and then cleaned with so-called Piranha etch solution (a mixture of H$_2$SO$_4$ and H$_2$O$_2$ at a ratio of 1:1) to remove organic matter attached to
the BiGa:LuIG surface \cite{Kikkawa2017PRB}. 
Subsequently, the Pt strip was sputtered on the BiGa:LuIG (100) surface by DC sputtering in an Ar atmosphere. 
The length and width of the Pt strip are $2~\textrm{mm}$ and $0.1~\textrm{mm}$, respectively. 
\par
To measure the SPE-induced temperature modulation, we follow the experimental method shown in Ref. \cite{Itoh2017PRB}. 
We first attached an E-type thermocouple (TC) on top of the middle of the Pt strip by pasting GE-varnish thinly [Fig. \ref{fig:1}(a)]. Hence, the TC wire is electrically insulated from but thermally well connected to the Pt layer. Then, the TC wires were connected to a nanovoltmeter (Keithley 2182A) via conductive wires, while the ends of the 2-mm-long Pt strip were connected to a current source (Keithley 6221).  
The magnetic field ${\bf H}$ (with magnitude $H$) was applied in the film plane and perpendicular
to the Pt strip, i.e., ${\bf H} \;||\; \mathbf{\hat{z}}$ in Fig. \ref{fig:1}(a), except for the control experiment shown in Fig. \ref{fig:3}(d), where ${\bf H} \;||\; \mathbf{\hat{x}}$.
For the electric detection of the SPE, it is important to exclude the large contribution from the Joule heating of the Pt layer \cite{Daimon2016NatCommun,Daimon2017PRB,Itoh2017PRB}. 
To this end, we applied a square-wave charge current $J_{\rm c}$ with amplitude $\Delta J_{\rm c}$ [Fig. \ref{fig:1}(b)] and measured the TC voltage $V_{\rm TC}$ that responds to the change in the $J_{\rm c}$ polarity, $\Delta V_{\rm TC} \equiv V_{\rm high}-V_{\rm low}$, where $V_{\rm high}$ ($V_{\rm low}$) represents the $V_{\rm TC}$ value when $J_{\rm c} = + \Delta J_{\rm c}$ $(- \Delta J_{\rm c})$  [see Fig. \ref{fig:1}(c)] \cite{Itoh2017PRB}.
Here, the Joule-heating-induced temperature modulation ($\propto \Delta J_{\rm c}^2$) takes a constant value in time [$V_{\rm offset}$ in Fig. \ref{fig:1}(c)] and does not overlap in $\Delta V_{\rm TC}$, enabling the electric detection of the SPE.  
The TC voltage $V_{\rm high}$ ($V_{\rm low}$) was recorded after the time delay $t_{\rm delay}$ of $10~\textrm{ms}$ \cite{Itoh2017PRB} and accumulated by repeating the process of the $J_{\rm c}$-polarity change from  $10^5$ to $10^6$ times for each $H$ point [see Fig. \ref{fig:1}(c)] to improve the signal-to-noise ratio and discern very tiny magnon-polaron signals (as small as sub-nanovolts) \cite{polarity-change}.   
$\Delta V_{\rm TC}$ can be converted into the corresponding temperature value $\Delta T ~(= \Delta V_{\rm TC}/S_{\rm TC})$ by the Seebeck coefficient of the E-type TC $S_{\rm TC}$. 
Further details on the measurements are shown in Sec. A of Supplemental Material (SM) \cite{SM}. \par 
%
%
\section{III.~~RESULTS AND DISCUSSION}
%
%
%
%
%
%
We first checked the appearance of the conventional SPE signals in the present Pt/BiGa:LuIG sample at the temperature of $T = 300~\textrm{K}$ and a low-$H$ range, $\left\vert \mu_0 H\right\vert <10~\textrm{mT}$. 
Figure \ref{fig:2}(a) shows the $H$ dependences of $\Delta V_{\rm TC}$ for several $\Delta J_{\rm c}$ values. 
With the application of $\Delta J_{\rm c}$, clear $\Delta V_{\rm TC}$ signals appear and the sign of $\Delta V_{\rm TC}$ changes when the ${\bf H}$ direction is reversed. 
Figure \ref{fig:2}(b) shows the $\Delta J_{\rm c}$-amplitude dependence of $\Delta V_{\rm TC}$ at $\mu_0 H = 8~\textrm{mT}$. The $\Delta V_{\rm TC}$ intensity is proportional to $\Delta J_{\rm c}$. 
These results confirm that $\Delta V_{\rm TC}$ is generated by the SPE;
the former result is attributed to the change of the SPE-induced heat-current ${\bf J}_{\rm q}$ direction with respect to the relative orientation between the spin accumulation ${\bm \mu}_{\rm s}$ of Pt and magnetization ${\bf M}$ of BiGa:LuIG [see Eq. (\ref{eq:SPE})], while the latter result is due to the fact that ${\bm \mu}_{\rm s}$, the driving force of the SPE, scales with  $J_{\rm c}$ [see Eq. (\ref{eq:spin-accumulation})]. \par
We then applied a relatively high magnetic field to the sample, $\left\vert \mu_0 H\right\vert <0.8~\textrm{T}$, and measured the $H$ dependence of $\Delta V_{\rm TC}$ with the intervals of $10~\textrm{mT}$. 
Figure \ref{fig:3}(b) displays the measured $\Delta V_{\rm TC}\left( H \right)$.
We observed a fine asymmetric peak structure at $\mu_0 H_{\rm peak} \sim \pm \; 0.4~\textrm{T}$ on top of the flat background signal.
A magnified view of the $\Delta V_{\rm TC}$-$H$ curve for $\Delta J_{\rm c} = 10.0~\textrm{mA}$ is shown in Fig. \ref{fig:3}(c), where the anomaly is marked by a blue triangle. 
The peak height $\Delta V_{\rm TC}^{\rm peak}$ is as small as $\sim 1.2~\textrm{nV}$, corresponding to the temperature change of $\Delta T_{\rm peak} \sim 0.02~\textrm{mK}$, but the structure is fully reproducible.  
We confirmed that the peak disappears when $J_{\rm c} = 0.0~\textrm{mA}$ [Fig. \ref{fig:3}(c)] and also when the $H$ is applied perpendicular to the film plane [${\bf H} \;||\; \mathbf{\hat{x}}$, see Fig. \ref{fig:3}(d)], consistent with the characteristics of the SPE [Eq. (\ref{eq:SPE})]. 
Besides, the peak amplitude $\Delta V_{\rm TC}^{\rm peak}$ was found to scale with $\Delta J_{\rm c}$ [see Fig. \ref{fig:3}(e)].
These results suggest that the anomaly is intrinsic and stems from the SPE. \par 
Importantly, the peak appears for the field $H_{\rm TA}$ at which the magnon dispersion curve of BiGa:LuIG touches the TA-phonon dispersion curve, as with the magnon-polaron anomaly observed in the SSE measurement \cite{Ramos2019NatCommun}. 
For BiGa:LuIG, the acoustic magnon dispersion reads $\omega \sim \gamma \mu_0 H + (J_{\rm ad}S/\hbar) A\{ [ (4/3)Ba^2k^2 + C^2 ]^{1/2} -C\}$ \cite{comment_on_magnon_dispersion}, where $\gamma$ is the gyromagnetic ratio, $\hbar$ the Dirac constant, $S$ the spin of Fe$^{3+}$ ($S=5/2$), $a$ ($J_{\rm ad}$) the nearest-neighbor distance (inter-sublattice exchange constant) between Fe a- and d-sites, and $A$, $B$, and $C$ the constants determined by the occupation ratio of Fe$^{3+}$ in the octahedral and tetrahedral sites of BiGa:LuIG \cite{Ramos2019NatCommun}. 
On the other hand, the phonon dispersions read $\omega = c_{\rm TA(LA)} k$, where the $c_{\rm TA(LA)}$ is the phonon sound velocity for the transverse-acoustic (TA) [longitudinal-acoustic (LA)] mode and determined as $c_{\rm TA(LA)} = 2.9~\textrm{km}/\textrm{s}$ ($6.2~\textrm{km}/\textrm{s}$) \cite{Ramos2019NatCommun}.
At a low $H \; (\sim 0)$, the magnon branch intersects the TA-phonon branch twice [see Fig. \ref{fig:3}(a)].
By increasing $H$, the magnon branch shifts toward high frequencies due to the Zeeman interaction ($\propto \mu_{0}H$), while the phonon branch remains unchanged. 
When $H$ becomes $\mu_0 H_{\rm TA} \sim 0.4~\textrm{T}$, the TA-phonon branch becomes tangential to the magnon dispersion [see Fig. \ref{fig:3}(a)]. 
For $H > H_{\rm TA}$ the branches are apart from each other.   
At the touching condition $H = H_{\rm TA}$, the largest overlap region between the magnon and phonon dispersion curves leads to the maximal magnon-polaron formation \cite{Kikkawa2016PRL,Flebus2017PRB}.  
Furthermore, the observed peak shape is close to that observed in the SSE measurement \cite{Ramos2019NatCommun}. 
These suggest that the observed peak is attributed to the magnon-polaron-induced SPE enhancement. 
The peak intensity relative to the background signal is $\sim 6~\%$, in reasonable agreement with that observed in the SSE ($\sim 10~\%$ \cite{Ramos2019NatCommun}). 
\par
%
%
%
\begin{figure}[htb]
\begin{center}
\includegraphics{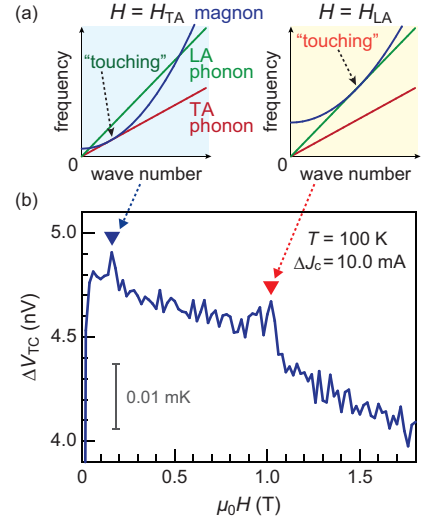}
\end{center}
\caption{(a) Schematics of the magnon, TA-phonon, and LA-phonon dispersion relations for BiGa:LuIG when $H=H_{\rm TA}$ and $H_{\rm LA}$.  
(b) $\Delta V_{\rm TC}\left( H \right)$ measured at $T=100~\textrm{K}$ and $\Delta J_{\rm c}=10~\textrm{mA}$ for $\mu_0 H <1.8~\textrm{T}$ [higher than that for Fig. \ref{fig:4}(a) to observe the hybridized magnon$-$LA-phonon feature]. 
The gray scale bar in (b) represents $\Delta T = 0.01~\textrm{mK}$. 
The $\Delta V_{\rm TC}$ peaks at $H_{\rm TA}$ and $H_{\rm LA}$ are marked by blue and red triangles, respectively.
The magnon-polaron signal at $H_{\rm TA}$ is clearer than that shown in Fig. \ref{fig:4}(a) (for 100 K). This stems from the improved signal-to-noise ratio for Fig. \ref{fig:5}(b) by the increased data accumulation time; the data was obtained by averaging 30-times repetition measurements [that are the highest accumulation number in the present work and took $\sim 7$ days for the data acquisition (note that the $H$ range is also increased compared to that for Fig. \ref{fig:4}(a)), see also Ref. \cite{polarity-change}].  
}
\label{fig:5}
\end{figure}
%
%
We carried out systematic $T$ dependence measurements. 
Figure \ref{fig:4}(a) shows the measured $\Delta V_{\rm TC}\left( H \right)$ for the various sample temperatures of $100~\textrm{K} \leq T \leq 325~\textrm{K}$ with $\Delta J_{\rm c} = 10.0~\textrm{mA}$.
At all the $T$ values, clear SPE signals were observed.
In Fig. \ref{fig:4}(b), we plot the $T$ dependences of $\Delta V_{\rm TC}$ (blue plots) and the corresponding $\Delta T$ values (gray plots) at $\mu_0 H =0.085~\textrm{T}~(< \mu_0 H_{\rm TA})$.  
When $T$ is decreased from $325~\textrm{K}$, the SPE-induced $\Delta T$ takes a broad maximum around $250~\textrm{K}$ and it monotonically decreases by further decreasing $T$. 
The $T$ dependence is qualitatively consistent with the previous results on the SSE in Pt/YIG films \cite{Kikkawa2015PRB,Jin2015PRB,Guo2016PRX,comment_on_T-dep_0}. 
Importantly, as marked by the blue triangles in Fig. \ref{fig:4}(a), magnon-polaron-induced peaks are also visible on top of the smooth background  at specific field values $H_{\rm peak}$.
The peak field $H_{\rm peak}$ shifts toward lower values by decreasing $T$ [see Figs. \ref{fig:4}(a) and \ref{fig:4}(d)]. 
This behavior is consistent with the magnon-polaron SSE in Pt/BiGa:LuIG films \cite{Ramos2019NatCommun} (see Sec. B of SM, where the previous magnon-polaron SSE peak fields are shown as a function of $T$ \cite{SM}) and attributable to the $T$ dependence of the stiffness parameter ($\propto J_{\rm ad}$) of the magnon dispersion \cite{Ramos2019NatCommun}. 
Besides, the magnon-polaron peak amplitude $\Delta V_{\rm TC}^{\rm peak}$, or $\Delta T_{\rm peak}$, tends to decrease with decreasing $T$ [Fig. \ref{fig:4}(c)], in agreement with the SSE results in Pt/BiGa:LuIG films in the same $T$ range \cite{Ramos2019NatCommun,comment_on_T-dep_1} (see Sec. B of SM, where the details on the determination of peak intensity and field are described \cite{SM}). \par 
We  found the magnon-polaron peak structure appears also when the magnon branch tangentially touches the LA-phonon dispersion. 
Figure \ref{fig:5}(b) shows the $H$ dependence of $\Delta V_{\rm TC}$ at $T = 100~\textrm{K}$ for $\Delta J_{\rm c} = 10.0~\textrm{mA}$ and $\mu_0 H < 1.8~\textrm{T}$  \cite{magnon-LA-phonon-touching}.  
Interestingly, in addition to the peak at $\mu_0H_{\rm TA} \sim 0.16~\textrm{T}$, a similar peak structure manifests at $\mu_0H_{\rm peak,2} \sim 1.0~\textrm{T}$ on top of the background signal, as marked by the blue and red triangles, respectively. 
Here, the overall negative slope in the background is due to the reduction of SPE by the field-induced freeze-out of the magnons \cite{Itoh2017PRB,Kikkawa2015PRB,Jin2015PRB,Guo2016PRX}. 
The peak-field value $H_{\rm peak,2}$ is again in reasonable agreement with those we have measured for the magnon-polaron SSE in Pt/BiGa:LuIG at the magnon$-$LA-phonon touching field $H_{\rm LA}$ [Fig. \ref{fig:5}(a)] \cite{Ramos2019NatCommun}, indicating that the peak at $H_{\rm peak,2}$ is originated from the magnon polarons induced by the coupling between magnons and LA-phonons in BiGa:LuIG. 
The result further corroborates that the SPE is enhanced by the magnon-polaron formation. \par
The appearance of the peaks in the SPE can be understood in terms of the enhanced lifetime of spin current by the magnon-phonon hybridization in a similar way as the SSE \cite{Kikkawa2016PRL,Flebus2017PRB,Ramos2019NatCommun}.
Assuming that the phonon lifetime is longer than the magnon one, the spin current carried by the hybridized modes exhibits a longer lifetime than pure magnonic spin currents \cite{Kikkawa2016PRL,Flebus2017PRB}. 
Hence, in the presence of the hybridization, the amount of spin current in BiGa:LuIG that reaches the interface and interacts with electrons of Pt can be increased, which reinforces the SPE. 
The effect is maximized when the magnetic field is tuned to the touching condition $H_{\rm TA (LA)}$, since the magnons and phonons can be hybridized over 
the largest volume in $k$-space \cite{Flebus2017PRB}), which shows up as the peak structures in the SPE signal at $H_{\rm TA (LA)}$.
\par
%
\section{IV.~~CONCLUSION}
%
To summarize, we observed anomalous peaks in $H$-dependent SPE signals in a Pt/BiGa:LuIG system. 
The anomalies appear at the onset $H$ values for hybridized magnon-phonon excitations (magnon-polaron formation), at which the magnon and phonon branches in BiGa:LuIG touch with each other. 
The observed peaks can be interpreted as the enhanced spin current caused by its lifetime increase due to the hybridization between magnons and long-lived phonons in BiGa:LuIG, which leads to the increased heat-current generation in the SPE. 
Future works should address the experimental validity of the Onsager reciprocity \cite{Flebus2017PRB,Ohnuma2017PRB,Sola2019SciRep} for the magnon-polaron peaks between the SPE and SSE. 
Furthermore, exploring the continuous BiGa:LuIG thickness dependence of the SPE via thermography \cite{Daimon2019} may be interesting to determine the length scales for magnon-polaron formation (magnon-phonon conversion) \cite{Rameshti2019PRB} and propagation \cite{Flebus2017PRB} from the interface.
\par 
%
\section*{ACKNOWLEDGMENTS}
%
%
We thank K. Uchida, R. Iguchi, H. Arisawa, and Y. Hashimoto for valuable discussions. 
This work is a part of the research program of ERATO ``Spin Quantum Rectification Project'' (No. JPMJER1402) from JST, the Grant-in-Aid for Scientific Research on Innovative Area ``Nano Spin Conversion Science'' (No. JP26103005), Grant-in-Aid for Scientific Research (S) (No. JP19H05600), and Grant-in-Aid for Research Activity Start-up (Nos. JP19K21031 and JP19K21035) from JSPS KAKENHI, JSPS Core-to-Core program ``International Research Center for New-Concept Spintronics Devices'', World Premier  International Research Center Initiative (WPI) from MEXT, Japan. 
K.O. and T.H. acknowledge support from GP-Spin at Tohoku University. \par
%
%

%
%
%
%
%
%
%
%
%
%
%
%
%
%
%
%

\begin{thebibliography}{99}
\bibitem{polarity-change}  
The $H$-dependent data (for each temperature $T$ and charge current $J_{\rm c}$) shown in the main text are obtained by averaging $6\sim30$-times repeated measurements. Here, at each magnetic field $H$ value for one measurement, the TC voltage $V_{\rm TC}$ was accumulated by repeating the process of the $J_{\rm c}$-polarity change of $4 \times 10^4$. Hence, the total data accumulation number for each $H$ point is from $2.4 \times 10^5$ to $1.2 \times 10^6$.
\bibitem{comment_on_T-dep_0} 
We note that the $T$ dependence of measured $\Delta T$ originates from not only the intrinsic $T$-dependent SPE but also various $T$-dependent factors, such as the thermal conductivity and heat capacity of the GE-varnish and interfacial thermal resistance between the thermocouple, GE-varnish, and sample, which should be considered for further quantitative discussions.
\bibitem{comment_on_T-dep_1}  
Although the origin of the observed $T$ dependence of $\Delta V_{\rm TC}^{\rm peak}$ [Fig. \ref{fig:4}(c)] remains to be clarified, it may be explained by the $T$ dependence of the lifetime difference between magnons and phonons; the magnon lifetime of BiGa:LuIG may be reduced at higher temperatures due to strong magnon-magnon scatterings (magnetic fluctuation), since the magnetic ordering temperature of  BiGa:LuIG is close to room temperature $\sim 400~\textrm{K}$. Therefore, the effect of magnon-polaron formation on the SPE signal can become more (less) prominent at the higher (lower) $T$ ranges, which may be responsible for the observed $T$ dependence of $\Delta V_{\rm TC}^{\rm peak}$ [Fig. \ref{fig:4}(c)].
\end{thebibliography}
\end{document}